\documentclass[english,a4paper]{article}
\usepackage[T1]{fontenc}
\usepackage[cp1250]{inputenc}
\usepackage{amssymb}
\usepackage{hyperref}
\usepackage{amsfonts}
\usepackage{graphicx}

\begin{document}
\title{Field-free molecular orientation by THz laser pulses at high temperature}

\author{M. Lapert\footnote{Institut
f\"ur Quanteninformationsverarbeitung, Universit\"at Ulm, D-89069
Ulm, GERMANY} and D. Sugny\footnote{Laboratoire Interdisciplinaire
Carnot de Bourgogne (ICB), UMR 5209 CNRS-Universit\'e de
Bourgogne, 9 Av. A. Savary, BP 47 870, F-21078 DIJON Cedex,
FRANCE, dominique.sugny@u-bourgogne.fr}}

\maketitle

\begin{abstract}
We investigate to which extend a THz laser pulse can be used to
produce field-free molecular orientation at high temperature. We
consider laser pulses that can be implemented with the state of
the art technology and we show that the efficiency of the control
scheme crucially depends on the parameters of the molecule. We
analyze the temperature effects on molecular dynamics and we
demonstrate that, for some molecules, a noticeable orientation can
be achieved at high temperature.
\end{abstract}

\section{Introduction}

Manipulating the molecular rotational degree of freedom remains a
goal of primary interest in photochemistry with applications
extending from chemical reactivity to nanoscale design
\cite{newrabitz,shapiro,rice}. In this framework, molecular
alignment and orientation constitute a well-established topic both
from the experimental and theoretical points of view
\cite{stapelfeldt,seideman}. While the alignment process is now
well understood in the adiabatic \cite{friedrich} or sudden regime
\cite{averbukh2} with recent extensions such as the molecular
classical rotation \cite{averbukh1}, the deflection of aligned
molecules \cite{averbukh3}, the planar alignment
\cite{sugnyplanar1,sugnyplanar2,hcn} or the analysis of the
dissipation effects \cite{seidemandiss,sugny1}, work remains to be
done in order to control and produce molecular orientation with a
sufficient high efficiency. On the theoretical side, several basic
mechanisms have been proposed, built on intuitive or optimal
control strategies \cite{salomon,lapert}. Among others, we can
cite the kick mechanism which consists in a sudden impact to the
molecule by a half-cycle pulse (HCP)
\cite{HCPexp,kick1,kick2,kick3,kick4,kick5}, its combination with
a laser field \cite{averbukhHCP1,averbukhHCP2,daems} or the
$(\omega-2\omega)$ scheme
\cite{KlingCO,tehini,zhang,tehini2,2colorchinois,multicolorchinois,twocolorchinois,coreehyper}.
Due to the efficiency of the first process based on its asymmetric
temporal shape, most of the theoretical works have focused on its
application \cite{arvieu,sugny7,sugny4,sugny3,HCPcoree}. However,
recent studies \cite{ortigoso,sugny6} have pointed out the
inherent experimental difficulties associated to the use of such
pulses, which are distorted when they propagate in free space as
well as through focusing optics \cite{HCPpropa}. This phenomenon,
due to the DC component of the field, makes thus problematic the
experimental implementation of these techniques in the control of
molecular rotation. In this framework, a fundamental question is
whether it is possible to orient linear molecules in the THz
regime by using only zero area laser pulses. These latter do not
contain DC field and are therefore free of these propagation
distortions. This problem has been recently addressed
theoretically
\cite{sugny6,THzchinois,THzchinois2,THzchinois3,chinese} and
experimentally \cite{orientationHCP}, but no systematic study of
the efficiency of this process has been done. Note that this
question is not trivial since the sudden impact approximation
predicts no post-pulse orientation in this regime
\cite{kick2,kick3,sugny6}. We present in this paper a complete
analysis of this control strategy by considering different linear
molecules. We establish under which conditions this process is
efficient and we analyze its robustness with respect to
temperature effects. Two mechanisms leading to molecular
orientation are identified. The first one is valid at low
temperature, while the second process is only efficient for higher
temperatures. In this second non-intuitive control scheme, we show
the positive role of temperature effects in the orientation
mechanism.

The paper is organized as follows. The model system is presented
in Sec. \ref{sec2}. The different numerical results are discussed
in Sec. \ref{sec3}. Conclusions and prospective views are given in
Sec. \ref{sec4}.
\section{The model system}\label{sec2}
We consider the control of a linear polar molecule by a linearly
polarized THz laser field $E(t)$ of zero area. The molecule is
assumed to be in its ground vibronic state. Within the rigid rotor
approximation, the Hamiltonian of the system can be written as
\begin{equation}
H(t)=BJ^2-\mu_0 E(t)\cos\theta,
\end{equation}
where $B$ is the rotational constant, $J^2$ the angular momentum
operator and $\mu_0$ the permanent dipolar moment. We neglect in
this paper the effect of the polarizability components since the
maximum intensity of the electric field remains moderate. The
units used throughout the paper are atomic units unless otherwise
specified. At non zero temperature, the system can be either
described by a density matrix $\rho(t)$ with a dynamics governed
by the von Neumann equation
\begin{equation}\label{eq1}
i\frac{\partial \rho(t)}{\partial t}=[H(t),\rho(t)],
\end{equation}
where $\rho(0)$ is the canonical density operator at thermal
equilibrium, or by a set of wave functions
$|\psi_{J_0,M_0}(t)\rangle$ satisfying each the Schr\"odinger
equation:
\begin{equation}\label{eq1b}
i\frac{\partial |\psi_{J_0,M_0}\rangle}{\partial
t}=H(t)|\psi_{J_0,M_0}\rangle,
\end{equation}
with as initial state $|\psi_{J_0,M_0}(t=0)\rangle
=|J_0,M_0\rangle$. The second representation given by Eq.
(\ref{eq1b}), which is more suited to the control mechanisms
interpretation, will be used in this work. The expectation value
$\langle \cos\theta\rangle$ defined by
\begin{equation}\label{eq2}
\langle \cos\theta\rangle
(t)=\frac{1}{Z}\sum_{J_0=0}^{+\infty}c_{J_0}\sum_{M_0=-J_0}^{M_0=J_0}\langle
\psi_{J_0,M_0}|\cos\theta |\psi_{J_0,M_0}\rangle,
\end{equation}
is taken as a quantitative measure of orientation, with the
weights $c_{J}=e^{-BJ(J+1)/(k_BT)}$ and the partition function
$Z=\sum_{J=0}^\infty\sum_{M=-J}^J c_J$, where $T$ is the
temperature and $k_B$ the Boltzman constant. In the rest of the
paper, we will need to separate the zero temperature contribution
to the thermal one, which are respectively denoted by $\langle
\cos\theta \rangle _0$ and $\langle \cos\theta \rangle _T$. In the
zero temperature response, the sum of Eq. (\ref{eq2}) is carried
out only for $J_0=0$, while for the thermal contribution, the
expectation value is computed over the other values of $J_0$.

The electric field is assumed to be of the form
\begin{eqnarray*}
E(t) &=& E_0f(t)=E_0\cos^2(\pi\frac{t}{\delta})\sin(2\pi ft),\quad t\in [ -\delta/2,\delta/2],\\
E(t) &=& 0\quad\textrm{otherwise},
\end{eqnarray*}
where $E_0$ is the amplitude of the pulse, $\delta$ its duration
and $f$ its central frequency. By symmetry, this field has a zero
area for any values of $\delta$ and $f$. Note that the sudden
impact approximation can be applied if the pulse duration $\delta$
is small with respect to the rotational period
$T_{\textrm{per}}=\pi/B$. This also means that noticeable
orientation can be produced only for sufficient large values of
$\delta$ \cite{kick2,sugny6}.

The Schr\"odinger equation (\ref{eq1b}) can be written as
\begin{equation}
i\frac{\partial |\psi_{J_0,M_0}\rangle}{\partial
\tau}=[J^2-A\cos\theta f(\tau)]|\psi_{J_0,M_0}\rangle,
\end{equation}
where the new dimensionless parameters are defined by:
\begin{eqnarray*}
\tau = Bt,~A = \frac{\mu_0E_0}{B},~F = \frac{f}{B},~D = B\delta,~
\tilde{T} = \frac{Tk_B}{B},
\end{eqnarray*}
with $f(\tau)=\cos^2(\pi\frac{\tau}{D})\sin(2\pi F\tau)$. In these
coordinates, the rotational period becomes $T_{\textrm{per}}=\pi$
and the pulse duration is $D$. These effective parameters
completely describe the dynamical evolution of the system and the
degree of molecular orientation produced. They will give a general
understanding of the orientation mechanism free of any particular
molecule. Note that the rotational constant $B$ is a crucial
parameter in these new coordinates since the effective field,
frequency and duration $(A,F,D)$ depend on $B$. For numerical
applications, we will consider the parameters listed in Table
\ref{tab1}.
\begin{table}\caption{\label{tab1}  Molecular parameters of different molecules used in the numerical computations. Numerical values are taken to be
$E_0=2.19$ MV/cm, $\delta=5$ ps and $f=0.5$ THz.}
\begin{center}
\begin{tabular}{c|c|c|c|c|c}
\hline
  Molecule& OCS &  HF  & LiH & CO & LiCl\\
  \hline \hline
  $B~(\textrm{cm}^{-1})$ &0.2029&20.956&7.513&1.931&1.345\\
\hline
 $ \mu_0~(\textrm{debye})$ &  0.712&1.820&5.88&0.112&6.33\\
\hline
 $A$&117.8497&2.9167&26.2842&1.9479&158.0563\\
\hline
$F$  &13.0823&0.1267&0.3533&1.3746&1.9735 \\
\hline $D$ & 0.1911&19.7371&7.0760&1.8187&1.2668   \\
\hline
 \end{tabular}
\end{center}
\end{table}
\section{Numerical Results}\label{sec3}
\begin{figure}[htbp]
\centering
\includegraphics[scale=0.6]{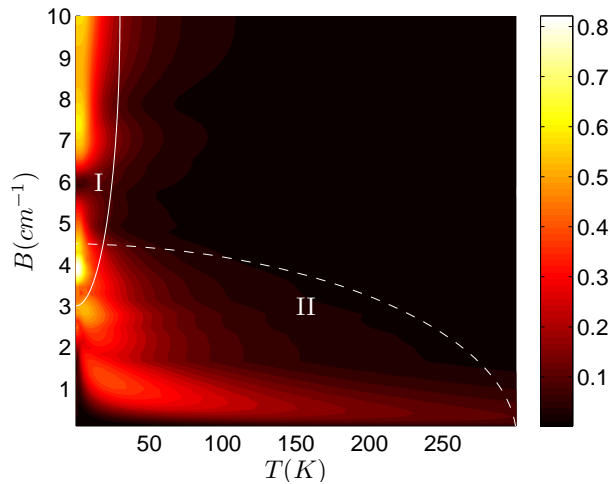}
\caption{(Color online) Maximum orientation as a function of the
rotational constant $B$ and the temperature $T$. The field
parameters are $\delta=5$ ps, $f=0.5$ THz and $E_0=2.19$ MV/cm.
Two regions (I) and (II) of the diagram are delimited by arcs of
ellipses. \label{fig1}}
\end{figure}

\begin{figure}[htbp]
{\centering\includegraphics[scale=0.6]{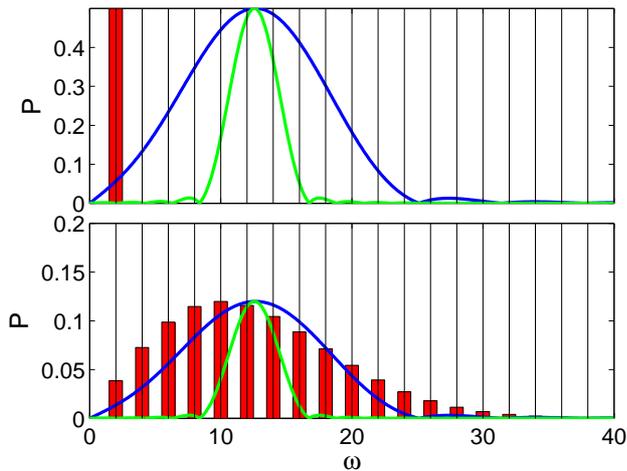}} \caption{(Color
online) Plot of the spectral distribution of the field at
$\tilde{T}=0$ and $\tilde{T}=50$ ($T=143.9$ K for $B=2$ cm$^{-1}$)
on the top and bottom panels, respectively. The amplitude of the
field spectrum is arbitrary. Control fields parameters are taken
to be $D=1$ ($\delta=2.654$ ps for $B=2$ cm$^{-1}$) for the blue
(black) line and $D=3$ ($\delta=7.963$ ps for $B=2$ cm$^{-1}$) for
the green (gray) line. Other parameters are fixed to $F=2$
($f=0.753$ THz for $B=2$ cm$^{-1}$) and $A=4$. The columns
represent the population distribution $P=(c_J+c_{J+1})/2$ at the
corresponding frequency $\omega=2(J+1)$. The quantities $P$ and
$\omega$ are unitless.\label{fig2}}
\end{figure}

\begin{figure}[htbp]
{\centering\includegraphics[scale=0.6]{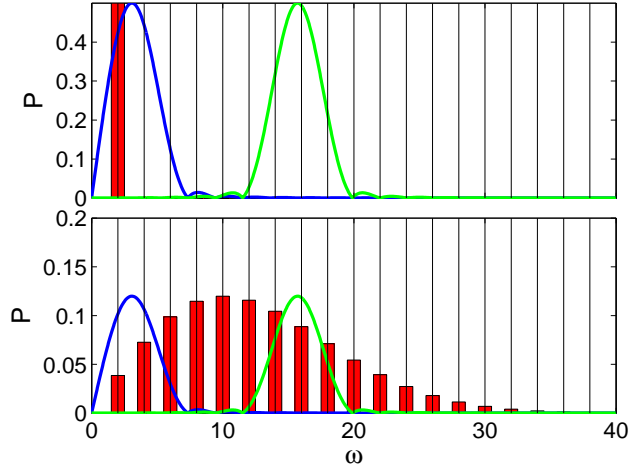}} \caption{(Color
online) Same as Fig. \ref{fig2} but for $F=0.5$ ($f_1=0.188$ THz
for $B=2$ cm$^{-1}$) in blue or dark and $F=2.5$ ($f=0.942$ THz
for $B=2$ cm$^{-1}$) in green or gray. The other parameters are
given by $D=3$ ($\delta=7.96$ ps for $B=2$ cm$^{-1}$) and
$A=4$.\label{fig3}}
\end{figure}

\begin{figure}[htbp]
\centering\includegraphics[scale=0.5]{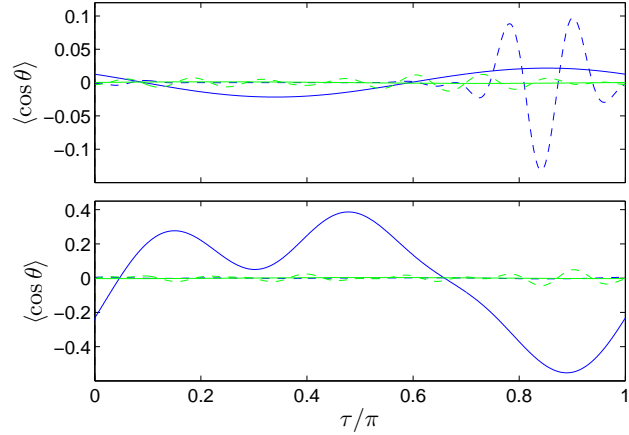} \caption{(Color
online) Time evolution of $\langle \cos\theta\rangle$ for the
cases of Fig. \ref{fig2} (top panel) and \ref{fig3} (bottom
panel). The same color code as in Fig. \ref{fig2} and \ref{fig3}
has been used. Solid and dashed lines depict respectively the
orientation at $\tilde{T}=0$ and $\tilde{T}=50$. The field is on
for negative times and switches off at $\tau=0$. The variable
$\tau$ is unitless.\label{fig4}}
\end{figure}

We begin our study by a general analysis of the maximum degree of
orientation that can be reached as a function of the rotational
constant $B$ and the temperature $T$. A fictive molecule with a
permanent dipolar moment of 1 debye has been considered in the
computation. The field amplitude is assumed to be $E_0=2.19$ MV/cm
(i.e. with a peak amplitude of 2 MV/cm), which corresponds roughly
to the maximum amplitude of THz pulses actually available
experimentally \cite{wolf}. This dependance is shown in Fig.
\ref{fig1} where two zones of high orientation can be clearly
distinguished: an upper zone, denoted (I), associated to high
values of $B$ and low temperature and a flat zone, denoted (II)
requiring small rotational constants and larger temperatures up to
250 K.
If the rotational constant is lower than 2 cm$^{-1}$, we first
notice that a very weak orientation is produced at $T=0$ K. The
rotational period varying as the inverse of $B$, this result can
be explained through the sudden impact approximation
\cite{kick2,kick3,sugny6}. Another standard feature shown in
previous studies is the detrimental effect of temperature on
molecular orientation \cite{seideman,kick2,kick3}. This behavior
can be recovered in the case of region (I) where almost no
orientation is obtained for a temperature larger than 100 K. By
comparison, the orientation observed in zone (II) is rather
unexpected, since the temperature effect becomes positive, no
orientation being produced at very low temperature. One of the
goal of this work will be to explore the basic mechanism at the
origin of this non trivial phenomenon.\\
\textbf{The orientation mechanism.}\\
We first present a general spectral analysis of the control
problem. We use in the following the dimensionless coordinates
introduced in Sec. \ref{sec2}, for which the molecular spectrum is
a discrete spectrum composed of the frequencies $\omega_J=2(J+1)$.
To simplify the discussion, we associate to each frequency a
weight $P$ which is defined as the average of the initial thermal
population of the levels $J$ and $J+1$, i.e. $P=(c_J+c_{J+1})/2$.
The spectrum of the control field is proportional to the one of
the function $f(\tau)$ and can roughly be approximated by a
gaussian spectrum centered at $\nu=F$ with a bandwidth
proportional to $1/D$. Different spectra are represented in Fig.
\ref{fig2} and \ref{fig3} for different values of $D$ and $F$,
respectively. In the case of Fig. \ref{fig2}, only the width of
the field spectrum is changed, while only the central frequency is
modified in Fig. \ref{fig3}. From a qualitative point of view, the
overlap between the molecular and the field spectra is a necessary
condition to produce molecular orientation. Figure \ref{fig2}
displays two cases with different values of $D$, i.e. a pulse with
a large spectrum overlapping many transitions including the first
one, and a pulse with a larger width $D$ corresponding to a
narrower spectrum. The first pulse can produce orientation at zero
and non-zero temperatures, while the second one is only efficient
at non zero temperature. This behavior is confirmed in the top
panel of Fig. \ref{fig4}. Figure \ref{fig3} depicts two different
situations, a first one with a low main frequency which overlaps
with the first frequency spectrum and a second case where the main
frequency is far from the first molecular transition. The first
pulse should work at zero and nonzero temperatures due to the
overlap with other transitions while in the second example, we
cannot expect orientation at zero temperature but only for
$\tilde{T}\neq 0$. The corresponding orientation responses are
given in the bottom panel of Fig. \ref{fig4}. Comparing in Fig.
\ref{fig4} the efficiency of the pulses of Fig. \ref{fig2} and
\ref{fig3}, one observes that at $\tilde{T}\neq 0$, a stronger
orientation is obtained in the first case of Fig. \ref{fig2} due
to a larger overlap of the field spectrum with the population
distribution. In other words, a noticeable orientation is produced
only for a large bandwidth of the laser field.
%In this situation,
%many rotational frequencies are simultaneously excited, the
%orientation being obtained by the constructive interferences of
%the different wave packets $|\psi_{J_0,M_0}(t)\rangle$.


The interpretation of the previous results leads to two basic
mechanisms governing the orientation response. The orientation
created at $T\simeq 0$ K can be associated to a rotational ladder
climbing mechanism from the ground energy level, which consists
here in the successive excitation of neighboring rotational
levels. This process is efficient only if the field spectrum
allows to excite the first rotational frequencies with a
sufficient high intensity. Note that this control scheme has
already been identified in the literature to produce molecular
orientation \cite{salomon}. The rotational population distribution
being displaced to high $J$ levels with increasing temperatures,
this mechanism then loses its efficiency very fast. This
orientation will be called a \emph{zero-temperature orientation}
and can be quantitatively measured by the partial expectation
value $\langle \cos\theta\rangle _0$. In the case of an initial
thermal distribution of rotational states, a new mechanism occurs.
This situation corresponds to molecules with small values of $B$,
i.e. with a quite large frequency $F$. As can be seen in Fig.
\ref{fig3} for $F=2.5$, there is no overlap with the population
distribution at low temperature and thus no orientation. In this
second mechanism at high temperature, the control field excites
simultaneously many rotational frequencies (i.e. with a large
laser bandwidth), creating several rotational waves packets
$|\psi_{J_0,M_0}(t)\rangle$ which interfere constructively to
produce a noticeable orientation, denoted \emph{thermal
orientation}. This latter can be computed through the expectation
value $\langle \cos\theta\rangle _T$. In this scenario, the
orientation reaches a maximum at a temperature different from zero
and presents a slow decrease with increasing temperatures.

\begin{figure}[htbp]
\centering\includegraphics[scale=0.6]{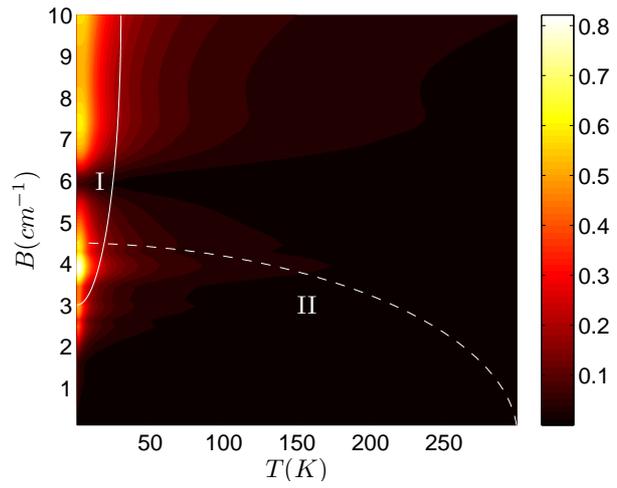}
\centering\includegraphics[scale=0.6]{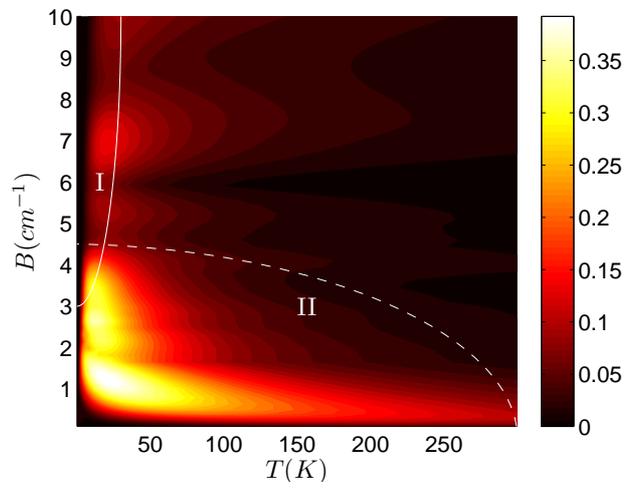} \caption{(Color
online) Same as Fig. \ref{fig1} but for the zero temperature
contribution $\langle \cos\theta\rangle_0$ to the molecular
orientation (top) and for the thermal one $\langle
\cos\theta\rangle_T$ (bottom).\label{fig5}}
\end{figure}

These different conclusions can be checked in Fig. \ref{fig5}
where the zero-temperature and the thermal orientation responses
have been plotted. One clearly sees in this figure that the zone
(I) is associated to $\langle \cos\theta\rangle _0$, where only
the initial state $|0,0\rangle$ is considered. The efficiency of
THz pulses in the region (II) can be interpreted as a thermal
orientation such that $\langle \cos\theta\rangle \simeq \langle
\cos\theta\rangle _T$. Note that the sum of the two figures
\ref{fig5} does not give exactly the result of Fig. \ref{fig1} due
to destructive interferences between the two orientation
responses.

The transition from a regime with thermal orientation to a regime
with a zero-temperature orientation can be understood from the
definition of the parameters $(F,D)$. Decreasing the value of $B$
is equivalent to increase the effective frequency $F$ and to
decrease the temporal width $D$ of the pulse. Starting from the
zone (I) where the spectrum of the field overlaps the first
molecular transition, a decrease of the rotational constant $B$
will shift the field spectrum to higher frequencies $F$. The
overlap with the first frequency is removed and no zero
temperature orientation is possible. At the same time, as a
consequence of the increase of the width of the field spectrum,
more and more molecular transitions can be excited by the pulse,
leading thus to a thermal orientation.

\begin{figure}[htbp]
\centering\includegraphics[scale=0.5]{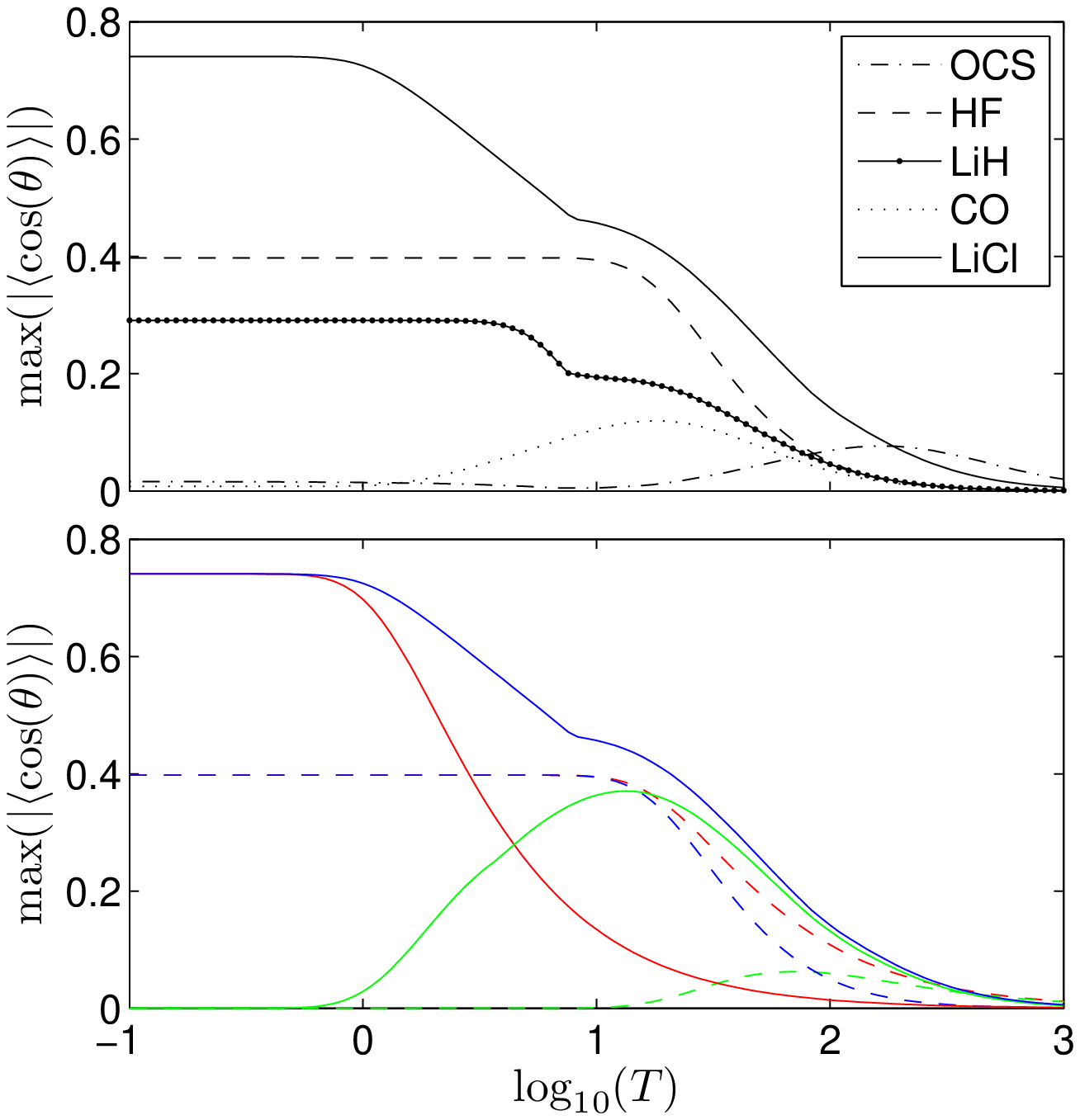} \caption{(Color
online) The top panel represents the evolution of the maximum of
orientation as a function of the temperature for the molecules of
Table \ref{tab1}. The field parameters of Fig. \ref{fig1} have
been used. The contribution of the zero temperature (red or dark
gray) and thermal (green of light gray) wave packets have been
plotted for the LiCl (solid lines) and HF (dashed lines) molecules
in the bottom panel. The blue (black) lines corresponds to the
total orientation response. The temperature $T$ is expressed in
Kelvin.\label{fig6}}
\end{figure}
The evolution of the orientation for the molecules of Table
\ref{tab1} is displayed in Fig. \ref{fig6}. All the molecules
considered belong to the zone (I), except for the CO and OCS
molecules which present an orientation close to 0 at $T\simeq 0 K$
and an orientation of the order of 0.15  and 0.1 at $T\simeq 10 K$
and $T\simeq 150 K$, respectively. The efficiency of the control
scheme at high temperature becomes even better than the one
obtained for other molecules such as HF or LiH, which is very good
only at low temperature. Note that the CO and LiCl molecules have
quite the same parameters $F$ and $D$. The difference of
orientation response comes from the difference of two order of
magnitude in their effective amplitude $A$. A ladder climbing
process at $T=0$ K is possible for LiCl since the tails of the
spectrum are sufficient, while it is not the case for CO. The two
molecules present however a thermal orientation. We also point out
that the orientation of OCS at high temperature is due to its low
value of $D$ (very broad spectrum) combined with a high value of
$F$ and a large effective amplitude of the field. In Fig.
\ref{fig6}, note the smooth evolution of the orientation, except
in the cases of the LiH and LiCl molecules where a slope change
occurs. This feature can be explained by a transition from a
zero-temperature orientation to a thermal one as shown in Fig.
\ref{fig6} (bottom). It can be seen that the non smooth point of
the curve for the LiCl molecule can be viewed as a limit point for
which the thermal orientation becomes predominant with respect to
the zero-temperature one.

\begin{figure}[htbp]
\centering\includegraphics[scale=0.5]{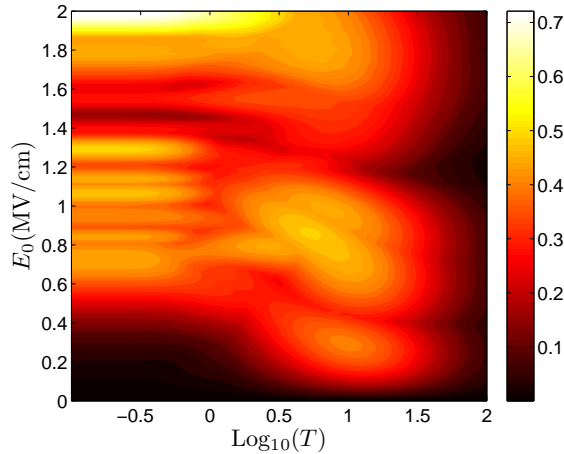} \caption{(Color
online) Maximum orientation as a function of the field amplitude
$E_0$ and the temperature $T$. The other field parameters are
$\delta=5$ ps and $f=0.5$ THz. The LiCl molecule is considered for
the computations. The temperature $T$ is expressed in
Kelvin.\label{fig7}}
\end{figure}
We finally explore the dependance of the orientation with respect
to the field strength. A global description of this sensitivity as
a function of the temperature is displayed in Fig. \ref{fig7}. At
zero temperature, we observe a quadratic increase of the
orientation up to $E_0\simeq 0.6$ MV/cm. Higher amplitudes lead to
a more chaotic evolution of the orientation characterized by the
occurrence of maxima and minima. At high temperature, the same
erratic distribution of the orientation response can be seen with,
however, a larger periodicity. We thus conclude that the molecular
orientation is less sensitive to amplitude field changes at
non-zero than at zero temperature. We also observe that the
thermal orientation can be produced at a lower intensity than the
zero temperature one.

%%%%%%%%%%%%%%%%%%%%%%%%%%%%%%%%%%%%%%%%%%%
\section{Conclusion}\label{sec4}
This paper has focused on the use of THz laser pulses for
controlling the orientation dynamics of linear molecules.
Numerical tests have shown the efficiency of the proposed control
scheme, even at high temperature for some molecules. Until now,
most of the works have envisaged the production of molecular
orientation with short laser pulses characterized by a non zero
time average of the electric field as, e.g., with the use of HCPs.
Indeed, the sudden impact approximation predicts no post-pulse
orientation when this average vanishes. Opposite to this accepted
fact, we have shown that a significant orientation can be obtained
in two different regimes. The first one corresponds to a standard
situation with low temperature and high rotational constants. In
the second case, for an adequate choice of the pulse parameters,
we have established that the temperature plays an active role in
the production of molecular orientation. This study calls for
further experimental investigation of the use of such laser pulses
in order to complete the initial work of Ref.
\cite{orientationHCP}. In particular, one objective could be to
demonstrate experimentally the existence of thermal orientation in
molecules such that OCS or CO.

\noindent \textbf{Acknowledgment}\\
We are grateful to E. Hertz for discussions.

%\bibliography{biblio1}
%\bibliographystyle{apsrev}

\end{document}